\input amstex
\documentstyle{amsppt}
%
\catcode`@=11
\redefine\output@{%
  \def\break{\penalty-\@M}\let\par\endgraf
  \ifodd\pageno\global\hoffset=105pt\else\global\hoffset=8pt\fi  
  \shipout\vbox{%
    \ifplain@
      \let\makeheadline\relax \let\makefootline\relax
    \else
      \iffirstpage@ \global\firstpage@false
        \let\rightheadline\frheadline
        \let\leftheadline\flheadline
      \else
        \ifrunheads@ 
        \else \let\makeheadline\relax
        \fi
      \fi
    \fi
    \makeheadline \pagebody \makefootline}%
  \advancepageno \ifnum\outputpenalty>-\@MM\else\dosupereject\fi
}
\catcode`\@=\active
\nopagenumbers
\def\negskp{\hskip -2pt}
\def\rot{\operatorname{rot}}
\def\const{\operatorname{const}}
\def\grad{\operatorname{grad}}
\accentedsymbol\ty{\tilde y}
\accentedsymbol\bry{\bar y}
\def\blue#1{#1}
\catcode`#=11\def\diez{#}\catcode`#=6
\catcode`_=11\def\podcherkivanie{_}\catcode`_=8
\def\mycite#1{\cite{\blue{#1}}}
\def\mytag#1{%
    \tag#1}
\def\mythetag#1{\thetag{\blue{#1}}}
\def\myrefno#1{\no#1}
\def\myhref#1#2{\blue{#2}}
\def\myEarXivlink{\myhref{http://arXiv.org}{http:/\negskp/arXiv.org}}
\pagewidth{360pt}
\pageheight{606pt}
\topmatter
\title
Gauge or not gauge\,?
\endtitle
\author
Ruslan Sharipov
\endauthor
\address Rabochaya street 5, 450003 Ufa, Russia
\endaddress
\email \vtop to 30pt{\hsize=280pt\noindent
\myhref{mailto:R\podcherkivanie Sharipov\@ic.bashedu.ru}
{R\_\hskip 1pt Sharipov\@ic.bashedu.ru}\newline
\myhref{mailto:r-sharipov\@mail.ru}
{r-sharipov\@mail.ru}\newline
\myhref{mailto:ra\podcherkivanie sharipov\@lycos.com}{ra\_\hskip 1pt
sharipov\@lycos.com}\vss}
\endemail
\urladdr
\myhref{http://www.geocities.com/r-sharipov}
{http:/\negskp/www.geocities.com/r-sharipov}
\endurladdr
\abstract
    The analogy of the nonlinear dislocation theory in crystals and 
the electromagnetism theory is studied. The nature of some quantities
is discussed. 
\endabstract
\endtopmatter
\loadbold
\TagsOnRight
\document

\head
1. Burgers vectors and the Burgers space.
\endhead
    Dislocations are one-dimensional defects of a crystalline grid
used to explain the plasticity in crystals. Each dislocation line is
characterized by its Burgers vector $\bold b$, while the dislocated 
medium in whole is characterized by the so-called {\it incompatible 
distorsion tensor $\hat\bold T$} (see \mycite{1}). 
\vadjust{\vskip 5pt\hbox to 0pt{\kern -10pt
\includegraphics{gauge01.eps}\hss}\vskip 155pt}The 
components of the Burgers vector for a dislocation line are determined
by the following path integral along some closed contour encircling 
this dislocation line (see Fig\.~1.1):
$$
\hskip -2em
b^{\,i}=\oint\limits_\gamma\sum^3_{j=1}\hat T^{\,i}_j\,dy^j.
\mytag{1.1}
$$
The Burgers vector $\bold b$ with components 
\mythetag{1.1} is teated as a vector of a special space, it is called
the {\it Burgers space}. The Burgers space is an imaginary space,
it is assumed to be filled with the infinite ideal (non-distorted)
crystalline grid. The tensor $\hat\bold T$ in \mythetag{1.1}
is a {\it double space tensor\/}: its upper index $i$ is associated
with some Cartesian coordinate system in the Burgers space, its 
lower index $j$ is a traditional tensorial index associated with
some coordinates $y^1, \,y^2,\,y^3$ (no matter Cartesian or 
curvilinear) in the real space where the crystalline medium moves.
\par
    In the continual limit, when the number of dislocation lines
is macroscopically essential, separate dislocation lines are replaced
by their distribution $\boldsymbol\rho$ (see \mycite{1} for more 
details). Then \mythetag{1.1} is replaced by the following integral
equality:
$$
\hskip -2em
\oint\limits_{\partial S}\sum^3_{j=1}\hat T^i_j\,dy^j=
\int\limits_S\sum^3_{j=1}\rho^{\,i}_j\,n^j\,dS.
\mytag{1.2}
$$
Here $S$ is some imaginary surface within the medium, $n^j$ are the
components of the unit normal vector to $S$, and $\gamma=
\partial S$ is the boundary of $S$ (see Fig\.~1.2). According to
\mythetag{1.2}, the double space tensorial quantity $\boldsymbol\rho$
is interpreted as the Burgers vector per unit area. Applying the Stokes
formula to \mythetag{1.2}, we get the differential equality
$$
\hskip -2em
\boldsymbol\rho=\rot\hat\bold T.
\mytag{1.3}
$$
Apart from \mythetag{1.3}, we have the following equality (see \mycite{1}):
$$
\hskip -2em
\bold j=-\grad\bold w-\frac{\partial\hat\bold T}{\partial t}.
\mytag{1.4}
$$
The double space tensorial quantity $\bold j$ is interpreted as the 
Burgers vector crossing the unit length of a contour per unit time
due to the moving dislocations (see \mycite{1}). However, the interpretation
of the quantity $\bold w$ was not clarified in \mycite{1}. This is the
goal of the present paper. For this purpose below we study two special
cases.
\head
2. Plastic relaxation.
\endhead
     Let's consider a two-dimensional model of a crystalline medium with square cells (see Fig\.~2.1). \vadjust{\vskip 5pt\hbox to 0pt{\kern -10pt
\includegraphics{gauge02.eps}\hss}\vskip 145pt}On the preliminary
stage the crystal was distorted as shown on Fig\.~2.2. This distorsion is
described by the following deformation map:
$$
\hskip -2em
\cases
x^1=x^1(y^1,y^2)=y^1-y^2,\\
x^2=x^2(y^1,y^2)=y^2.
\endcases
\mytag{2.1}
$$
Here we assume that $x^1,\,x^2$ are Cartesian coordinates in the Burgers
space and $y^1,\,y^2$ are Cartesian coordinates in the real space, both
are associated with the orthonormal bases $\bold e_1,\,\bold e_2$ and 
$\bold E_1,\,\bold E_2$ respectively. By differentiating \mythetag{2.1} we
find the components of the {\it compatible distortion tensor} $\bold T$
(see \mycite{1}):
$$
\hskip -2em
T^i_k=\frac{\partial x^i}{\partial y^k}=\Vmatrix 1 & -1\\
0 & 1\endVmatrix.
\mytag{2.2}
$$
They are constants since the further evolution of our crystal goes
without the displacement of atoms (see Fig\.~2.3, Fig\.~2.4, Fig\.~2.5,
Fig\.~2.6). Hence, we have
$$
\hskip -2em
\frac{\partial\bold T}{\partial t}=0.
\mytag{2.3}
$$
Initially, our crystal has no dislocations at all. Therefore, the compatible
and incompatible distorsion tensors are initially equal to each other:
$$
\hskip -2em
\hat\bold T\,\vbox{\hrule width 0.5pt
height 8pt depth 8pt}_{\,\,t=t_0}=\bold T\,\vbox{\hrule width 0.5pt
height 8pt depth 8pt}_{\,\,t=t_0}.
\mytag{2.4}
$$\par
     On Fig\.~2.3 three pairs of the edge dislocations arise. Their Burgers
vectors are $\bold b=\bold e_1$ and $\bold b=-\bold e_1$ in each pair.
Therefore, the total Burgers vector \vadjust{\vskip 5pt\hbox to 0pt{\kern
-10pt\includegraphics{gauge03.eps}\hss}\vskip 110pt}of this
group of dislocations is equal to zero. On Fig\.~2.4, Fig\.~2.5, Fig\.~2.6
the dislocations with negative Burgers vectors $\bold b=-\bold e_1$ move to
the right. During the evolution time the blue arrow of the length $3$
(see Fig\.~2.3 and Fig\.~2.6) is crossed by three dislocations with total
Burgers vector $\bold b=-3\,\bold e_1$. The green arrow of the same length
is not crossed by the moving dislocations at all. Therefore, we have the equality
$$
\hskip -2em
\int\limits^{\ \,t_1}_{t_0} j^{\,i}_{\,k}\,dt=
\Vmatrix 0 & -1\\
0 & 0\endVmatrix
\mytag{2.5}
$$
at the center of our crystal. Behind the moving dislocations we find the
undistorted cells, they are marked by yellow spots on Fig\.~2.6. This means that
$$
\hskip -2em
\hat T^i_k\,\vbox{\hrule width 0.5pt
height 8pt depth 8pt}_{\,\,t=t_1}
=\Vmatrix 1 & 0\\ 0 & 1\endVmatrix
\mytag{2.6}
$$
at the center of our crystal. Combining \mythetag{2.2}, \mythetag{2.3},
\mythetag{2.4}, \mythetag{2.5}, and \mythetag{2.6}, we get
$$
\pagebreak
\hskip -2em
\hat\bold T\,\vbox{\hrule width 0.5pt height 14pt depth 12pt}
\vbox{\hrule width 0pt height 12pt depth 12pt}^{\ t_1}_{\ t_0}
+\int\limits^{\ \,t_1}_{t_0}\bold j\ dt=0.
\mytag{2.7}
$$
In the continuous limit, when the moving dislocations 
form the homogeneous and constant flow the above equality \mythetag{2.7} 
can be transformed to the following one:
$$
\hskip -2em
\frac{\partial\hat\bold T}{\partial t}+\bold j=0.
\mytag{2.8}
$$
Comparing \mythetag{2.8} with \mythetag{1.4}, we conclude that $\bold v=0$
and $\bold T=\const$ implies $\bold w=0$ in our first example.
\head
3. Frozen dislocations.
\endhead
     As the second example, we consider a three-dimensional crystal 
where the dislocation lines move together with the medium like 
water-plants frozen \vadjust{\vskip 5pt\hbox to 0pt{\kern
-10pt\includegraphics{gauge04.eps}\hss}\vskip 225pt}into 
the ice. The choice of Cartesian coordinates $x^1,\,x^2,\,x^3$ in the 
Burgers space is obligatory (see \mycite{1}). In the real space we could
choose either Cartesian or curvilinear coordinates. Below we choose
Cartesian coordinates $y^1,\,y^2,\,y^3$ for the sake of simplicity. Then 
the interspace map and its inverse map are given by the following formulas:
$$
\xalignat 2
&\hskip -2em
\cases\ty^1=x^1,\\ \ty^2=x^2,\\ \ty^3=x^3,\endcases
&&\cases x^1=\ty^1,\\ x^2=\ty^2,\\ x^3=\ty^3.\endcases
\mytag{3.1}
\endxalignat
$$
The evolution of the crystal is subdivided into two stages (see Fig\.~3.1).
In the first stage, which is a preliminary one, the dislocations are
produced: 
$$
\xalignat 2
&\hskip -2em
\cases\bry^1=\bry^1(\tau,\ty^1,\ty^2,\ty^3),\\
\bry^2=\bry^2(\tau,\ty^1,\ty^2,\ty^3),\\
\bry^3=\bry^3(\tau,\ty^1,\ty^2,\ty^3),\endcases
&&\cases 
\ty^1=\ty^1(\tau,\bry^1,\bry^2,\bry^3),\\
\ty^2=\ty^2(\tau,\bry^1,\bry^2,\bry^3),\\
\ty^3=\ty^3(\tau,\bry^1,\bry^2,\bry^3).\endcases
\mytag{3.2}
\endxalignat
$$
In the second stage the dislocations are frozen and move together 
with the medium:
$$
\xalignat 2
&\hskip -2em
\cases y^1=y^1(t,\tau,\bry^1,\bry^2,\bry^3),\\
y^2=y^2(t,\tau,\bry^1,\bry^2,\bry^3),\\
y^3=y^3(t,\tau,\bry^1,\bry^2,\bry^3),\endcases
&&\cases 
\bry^1=\bry^1(t,\tau,y^1,y^2,y^3),\\
\bry^2=\bry^2(t,\tau,y^1,y^2,y^3),\\
\bry^3=\bry^3(t,\tau,y^1,y^2,y^3).\endcases
\mytag{3.3}
\endxalignat
$$
The compatible distortion tensor $\bold T$ at the time instant $t$
is determined by the composite map including all of the three above
maps \mythetag{3.1}, \mythetag{3.2}, and \mythetag{3.3}:
$$
\hskip -2em
T^i_k(t,y^1,y^2,y^3)=\frac{\partial x^i}{\partial y^k}
\mytag{3.4}
$$
(see \mythetag{2.2} for comparison). Similarly, at the time instant $t=\tau$
we have
$$
\hskip -2em
T^i_q(\tau,\bry^1,\bry^2,\bry^3)=\frac{\partial x^i}{\partial\bry^q}.
\mytag{3.5}
$$
From \mythetag{3.4} and \mythetag{3.5}, applying the chain rule to
\mythetag{3.3}, we derive the relationship
$$
\hskip -2em
T^i_k(t,y^1,y^2,y^3)=\sum^3_{q=1}T^i_q(\tau,\bry^1,\bry^2,\bry^3)\,
\frac{\partial\bry^q}{\partial y^k}.
\mytag{3.6}
$$
The relationship \mythetag{3.6} expresses the evolution rule for the
compatible distortion tensor $\bold T$. If the dislocations are 
frozen into the material, then the incompatible distortion tensor 
$\hat\bold T$ should obey the same evolution rule:
$$
\hskip -2em
\hat T^i_k(t,y^1,y^2,y^3)=\sum^3_{q=1}\hat T^i_q(\tau,\bry^1,\bry^2,
\bry^3)\,\frac{\partial\bry^q}{\partial y^k}.
\mytag{3.7}
$$
For the sake of simplicity, in the further calculations we denote
$$
\hskip -2em
\bar T^q_k(t,\tau,y^1,y^2,y^3)=\frac{\partial\bry^q}{\partial y^k}.
\mytag{3.8}
$$
The quantities \mythetag{3.8} form a non-degenerate square matrix
$\bar T$. Let $\bar S=\bar T^{-1}$ be the inverse matrix for $\bar T$
and let $S^q_k=S^q_k(t,\tau,y^1,y^2,y^3)$ be the components of this
inverse matrix. Note that $\bar T^q_k$ are not the components of a 
tensor field. They are not the components of a double space tensor 
in the sense of \mycite{1} as well. We shall not discuss the tensorial
properties of the quantities $\bar T^q_k$, we shall use them only as 
the notations for the partial derivatives \mythetag{3.8}.\par
     The following functional identity with two parameters $t$ and 
$\tau$ is quite obvious for the pair of mutually inverse maps given 
by the functions \mythetag{3.3}:
$$
\hskip -2em
y^k(t,\tau,\bry^1(t,\tau,y^1,\,\ldots,y^3),\,\ldots,
\bry^3(t,\tau,y^1,\,\ldots,y^3)=y^k.
\mytag{3.9}
$$
By differentiating \mythetag{3.9} with respect to the time variable $t$
we easily derive
$$
\hskip -2em
\frac{\partial\bry^q}{\partial t}=-\sum^3_{r=1}v^r\,\bar T^q_r.
\mytag{3.10}
$$
Here $v^1,\,v^2,\,v^3$ are the components of the velocity vector 
$\bold v=\bold v(t,y^1,y^2,y^3)$ of a point of the medium. Applying
the partial derivative $\partial/\partial y^k$ to \mythetag{3.10} and
taking into account \mythetag{3.8}, we derive another useful equality:
$$
\hskip -2em
\frac{\partial\bar T^q_k}{\partial t}=-\sum^3_{r=1}
\frac{\partial(v^r\,\bar T^q_r)}{\partial y^k}.
\mytag{3.11}
$$
As for the equalities \mythetag{3.6} and \mythetag{3.7}, now they are
written as follows:
$$
\align
&\hskip -2em
T^i_k=\sum^3_{q=1}T^i_q(\tau,\bry^1,\bry^2,\bry^3)\ \bar T^q_k,
\\
&\hskip -2em
\hat T^i_k=\sum^3_{q=1}\hat T^i_q(\tau,\bry^1,\bry^2,\bry^3)\ 
\bar T^q_k.
\mytag{3.12}
\endalign
$$
\proclaim{Theorem 3.1} In the case of frozen dislocations the
density of the Burgers vector $\boldsymbol\rho$ obeys the following
evolution rule:
$$
\hskip -2em
\rho^i_k=\det\bar T\ \sum^3_{p=1}\sum^3_{q=1}\sum^3_{r=1}
\rho^i_q(\tau,\bry^1,\bry^2,\bry^3)\,g^{qr}\,\bar S^p_r
\,g_{pk}.
\mytag{3.13}
$$
Here $g_{pk}$ and $g^{qr}$ are the components of the metric tensor 
and the dual metric tensor respectively (see \mycite{2} for details).
\endproclaim
\demo{Proof} The proof is pure calculations. The density of the 
Burgers vector is defined by the formula \mythetag{1.3}. In coordinate
form this formula is written as
$$
\hskip -2em
\rho^i_k=\sum^3_{r=1}\sum^3_{p=1}\sum^3_{q=1}g_{kr}\ \omega^{rpq}\
\frac{\partial\hat T^i_q}{\partial y^p},
\mytag{3.14}
$$
where $\omega^{rpq}$ are the components of the so-called {\it volume
tensor} (see \mycite{2}). The quantities $\omega^{rpq}$ in \mythetag{3.14}
and the quantities $g^{qr}$ and $g_{pk}$ in \mythetag{3.13} are constants
because we chose the Cartesian coordinates $y^1,\,y^2,\,y^3$ 
(see Fig\.~3.1). From \mythetag{3.12} we derive
$$
\frac{\partial\hat T^i_q}{\partial y^p}=
\sum^3_{m=1}\frac{\partial\hat T^i_m(\tau,\bry^1,\bry^2,\bry^3)}
{\partial y^p}\ \bar T^m_q+\sum^3_{m=1}\hat T^i_m(\tau,\bry^1,\bry^2,
\bry^3)\ \frac{\partial\bar T^m_q}{\partial y^p}.
\qquad
\mytag{3.15}
$$
Then we apply the chain rule to the first term and the formula 
\mythetag{3.8} to the second term in the right hand side of 
the equality \mythetag{3.15}. As a result we get
$$
\frac{\partial\hat T^i_q}{\partial y^p}=\sum^3_{m=1}\sum^3_{n=1}
\frac{\partial\hat T^i_m(\tau,\bry^1,\bry^2,\bry^3)}
{\partial\bry^n}\ \frac{\partial\bry^n}{\partial y^p}\ \bar T^m_q
+\sum^3_{m=1}\hat T^i_m(\tau,\bry^1,\bry^2,
\bry^3)\ \frac{\partial^2\bry^m}{\partial y^p\,\partial y^q}.
$$
Applying the formula \mythetag{3.8} again, we derive
$$
\frac{\partial\hat T^i_q}{\partial y^p}=\sum^3_{m=1}\sum^3_{n=1}
\frac{\partial\hat T^i_m(\tau,\bry^1,\bry^2,\bry^3)}
{\partial\bry^n}\ \bar T^n_p\ \bar T^m_q
+\sum^3_{m=1}\hat T^i_m(\tau,\bry^1,\bry^2,
\bry^3)\ \frac{\partial^2\bry^m}{\partial y^p\,\partial y^q}.
$$
Now we substitute the above equality into \mythetag{3.14}. The second term 
in its right hand side is symmetric in $p$ and $q$. It vanishes 
when substituted into \mythetag{3.14} because of the skew symmetry of $\omega^{rpq}$ (see \mycite{2}). For $\rho^i_k$ now we get
$$
\hskip -2em
\rho^i_k=\sum^3_{r=1}\sum^3_{p=1}\sum^3_{q=1}\sum^3_{m=1}\sum^3_{n=1}
g_{kr}\ \omega^{rpq}\ \frac{\partial\hat T^i_m(\tau,\bry^1,\bry^2,
\bry^3)}{\partial\bry^n}\ \bar T^n_p\ \bar T^m_q.
\mytag{3.16}
$$
In the next step we use the following well-known identity:
$$
\det\bar T\cdot\,\omega^{\,lnm}=\sum^3_{r=1}\sum^3_{p=1}\sum^3_{q=1}
\omega^{rpq}\ \bar T^l_r\ \bar T^n_p\ \bar T^m_q.
$$
Since $\bar S=\bar T^{-1}$, from this identity we immediately derive
$$
\hskip -2em
\det\bar T\ \sum^3_{l=1}\,\bar S^r_l\ \omega^{\,lnm}
=\sum^3_{p=1}\sum^3_{q=1}\omega^{rpq}\ \bar T^n_p\ \bar T^m_q.
\mytag{3.17}
$$
Comparing \mythetag{3.16} and \mythetag{3.17} we can write the following
formula for $\rho^i_k$:
$$
\hskip -2em
\rho^i_k=\det\bar T\ \sum^3_{r=1}\sum^3_{l=1}\sum^3_{m=1}\sum^3_{n=1}
g_{kr}\ \bar S^r_l\ \omega^{\,lnm}\ \frac{\partial\hat T^i_m(\tau,
\bry^1,\bry^2,\bry^3)}{\partial\bry^n}.
\mytag{3.18}
$$
Now let's rewrite the formula \mythetag{3.14} for the time instant
$t=\tau$:
$$
\hskip -2em
\rho^i_q(\tau,\bry^1,\bry^2,\bry^3)
=\sum^3_{l=1}\sum^3_{n=1}\sum^3_{m=1}g_{ql}\ \omega^{\,lnm}\
\frac{\partial\hat T^i_m(\tau,\bry^1,\bry^2,\bry^3)}
{\partial\bry^n}.
\mytag{3.19}
$$
Comparing \mythetag{3.18} and \mythetag{3.19} we can rewrite \mythetag{3.18}
as follows:
$$
\hskip -2em
\rho^i_k=\det\bar T\ \sum^3_{r=1}\sum^3_{l=1}\sum^3_{q=1}
g_{kr}\ \bar S^r_l\ g^{lq}\ \rho^i_q(\tau,\bry^1,\bry^2,\bry^3).
\mytag{3.20}
$$
Now it is easy to see that, in essential, \mythetag{3.20} coincides with 
the equality  \mythetag{3.13} which we needed to prove. So the proof is
over.\qed\enddemo
\parshape 15 0pt 360pt 0pt 360pt 0pt 360pt
175pt 185pt 175pt 185pt 175pt 185pt 175pt 185pt 
175pt 185pt 175pt 185pt 175pt 185pt 175pt 185pt 175pt 185pt 
175pt 185pt 175pt 185pt 0pt 360pt
    In the case of frozen dislocations \vadjust{\vskip 5pt\hbox to 
0pt{\kern -20pt\includegraphics{gauge05.eps}\hss}\vskip
-5pt}the motion of the dislocation lines is completely determined by
the motion of the medium. Therefore, $\bold j$ should be expressed
through $\bold v$ and $\boldsymbol\rho$. In order to find this 
expression let's remember that $\bold j$ by definition is the total 
Burgers vector of the moving dislocations that cross the unit length 
of a contour $\gamma$ per unit time. It is clear that all of the
dislocations passing through the dark parallelogram on Fig\.~3.2 will 
cross the segment $\boldsymbol\tau$ during the next time interval $dt$.
The total Burgers vector of such dislocations is determined by formula
$$
db^i=\sum^3_{k=1}\rho^i_k\,n^k\,dS,\quad
\mytag{3.21}
$$
where $\bold n\,dS=[\bold v,\,\boldsymbol\tau]\,dt$ and $[\bold v,
\,\boldsymbol\tau]$ is the cross product (vector product) of $\bold v$ 
and $\boldsymbol\tau$. On the other hand, the same Burgers vector
is given by another formula
$$
\hskip -2em
db^i=\sum^3_{c=1}j^{\,i}_{\,c}\,\tau^c\,dt.
\mytag{3.22}
$$
If we write the equality $\bold n\,dS=[\bold v,\,\boldsymbol\tau]\,dt$ 
in coordinate form 
$$
n^k\,dS=\sum^3_{s=1}\sum^3_{l=1}\sum^3_{c=1}g^{ks}\,\omega_{slc}\
v^{\,l}\,\tau^c\,dt,
$$
then from \mythetag{3.21} and \mythetag{3.22} we derive the following formula
for $\bold j$:
$$
\hskip -2em
j^{\,i}_{\,c}=\sum^3_{s=1}\sum^3_{l=1}\sum^3_{k=1}\omega_{slc}\
v^{\,l}\,g^{sk}\,\rho^i_k.
\mytag{3.23}
$$
\proclaim{Theorem 3.2}In the case of frozen dislocation $\boldsymbol\rho$
and $\bold j$ are related to each other by the formula \mythetag{3.23}.
\endproclaim
    In order to continue the above calculations, now we substitute
\mythetag{3.16} into \mythetag{3.23}. As a result for the components of
$\bold j$ we derive the following expression:
$$
j^{\,i}_{\,c}=\sum^3_{s=1}\sum^3_{l=1}\sum^3_{p=1}\sum^3_{q=1}\sum^3_{m=1}\sum^3_{n=1}
\omega_{slc}\
v^{\,l}\ \omega^{spq}\ \frac{\partial\hat T^i_m(\tau,\bry^1,\bry^2,
\bry^3)}{\partial\bry^n}\ \bar T^n_p\ \bar T^m_q.
$$
Applying the well-known identity\footnotemark 
$$
\sum^3_{s=1}\omega_{slc}\ \omega^{spq}=
\delta^p_l\,\delta^q_c-\delta^p_c\,\delta^q_l
$$
\footnotetext{\ Here $\delta^p_l$, $\delta^q_c$, $\delta^p_c$,
and $\delta^q_l$ are the Kronecker symbols.}
to this expression we can decrease the multiplicity of summation
symbols in it:
$$
\aligned
j^{\,i}_{\,c}=&\sum^3_{p=1}\sum^3_{m=1}\sum^3_{n=1}\bar T^m_c\ v^{\,p}\ 
\bar T^n_p\ \frac{\partial\hat T^i_m(\tau,\bry^1,\bry^2,\bry^3)}
{\partial\bry^n}-\\
&-\sum^3_{q=1}\sum^3_{m=1}\sum^3_{n=1}\bar T^n_c\ v^{\,q}\ 
\bar T^m_q\ \frac{\partial\hat T^i_m(\tau,\bry^1,\bry^2,\bry^3)}
{\partial\bry^n}.
\endaligned
\mytag{3.24}
$$
\proclaim{Theorem 3.3}In the case of frozen dislocation $\bold j$
and $\hat\bold T$ are related to each other by the formula \mythetag{3.24}.
\endproclaim
    Now we are going to calculate the time derivative of $\hat\bold T$ 
using \mythetag{3.12}. Upon doing this, we will be able to verify the
equality \mythetag{1.4} and calculate $\bold w$:
$$
\gather
\frac{\partial\hat T^i_c}{\partial t}=
\sum^3_{q=1}\frac{\partial\hat T^i_q(\tau,\bry^1,\bry^2,\bry^3)}
{\partial t}\ \bar T^q_c+\sum^3_{q=1}\hat T^i_q(\tau,\bry^1,\bry^2,
\bry^3)\ \frac{\partial\bar T^q_c}{\partial t}=\\
\displaybreak
=\sum^3_{q=1}\sum^3_{r=1}
\frac{\partial\hat T^i_q(\tau,\bry^1,\bry^2,\bry^3)}{\partial\bry^r}
\,\frac{\partial\bry^r}{\partial t}\ \bar T^q_c-\sum^3_{q=1}\sum^3_{p=1}
\hat T^i_q(\tau,\bry^1,\bry^2,\bry^3)\ \frac{\partial(v^p\,
\bar T^q_p)}{\partial y^c}=\\
=-\sum^3_{q=1}\sum^3_{r=1}\sum^3_{p=1}
\frac{\partial\hat T^i_q(\tau,\bry^1,\bry^2,\bry^3)}{\partial\bry^r}
\ v^p\ \bar T^r_p\ \bar T^q_c-\sum^3_{q=1}\sum^3_{p=1}
\hat T^i_q(\tau,\bry^1,\bry^2,\bry^3)\ \frac{\partial(v^p\,
\bar T^q_p)}{\partial y^c}.
\endgather
$$
In the above calculations we used the identities \mythetag{3.10} and
\mythetag{3.11}. In order to make the resulting expression similar to
\mythetag{3.24} we change some summation indices:
$$
\gathered
\frac{\partial\hat T^i_c}{\partial t}=-\sum^3_{p=1}\sum^3_{m=1}
\sum^3_{n=1}\bar T^m_c\ v^p\ \bar T^n_p\ \frac{\partial\hat T^i_m(\tau,
\bry^1,\bry^2,\bry^3)}{\partial\bry^n}\ -\\
-\sum^3_{m=1}\sum^3_{q=1}
\hat T^i_m(\tau,\bry^1,\bry^2,\bry^3)\ \frac{\partial v^q}{\partial y^c}
\ \bar T^m_q
-\sum^3_{m=1}\sum^3_{q=1}
\hat T^i_m(\tau,\bry^1,\bry^2,\bry^3)\ v^q\ \frac{\partial
\bar T^m_q}{\partial y^c}.
\endgathered\quad
\mytag{3.25}
$$
Now let's return to the equation \mythetag{1.4}. In coordinate form
it looks like
$$
\hskip -2em
\frac{\partial\hat T^i_c}{\partial t}+j^{\,i}_{\,c}
=-\frac{\partial w^i}{\partial y^c}.
\mytag{3.26}
$$
Substituting \mythetag{3.24} and \mythetag{3.25} into \mythetag{3.26}, we
derive
$$
\gathered
\frac{\partial w^i}{\partial y^c}=
\sum^3_{q=1}\sum^3_{m=1}\sum^3_{n=1}\bar T^n_c\ v^{\,q}\ 
\bar T^m_q\ \frac{\partial\hat T^i_m(\tau,\bry^1,\bry^2,\bry^3)}
{\partial\bry^n}+\\
+\sum^3_{m=1}\sum^3_{q=1}
\hat T^i_m(\tau,\bry^1,\bry^2,\bry^3)\ \frac{\partial v^q}{\partial y^c}
\ \bar T^m_q
+\sum^3_{m=1}\sum^3_{q=1}
\hat T^i_m(\tau,\bry^1,\bry^2,\bry^3)\ v^q\ \frac{\partial
\bar T^m_q}{\partial y^c}.
\endgathered\quad
\mytag{3.27}
$$
Let's remember the formula \mythetag{3.8} and apply it to $\bar T^n_c$
in \mythetag{3.27}:
$$
\gathered
\frac{\partial w^i}{\partial y^c}=\sum^3_{m=1}\sum^3_{q=1}
\left(\,\shave{\sum^3_{n=1}}\frac{\partial
\hat T^i_m(\tau,\bry^1,\bry^2,\bry^3)}{\partial\bry^n}\
\frac{\partial\bry^n}{\partial y^c}\right)v^{\,q}\ \bar T^m_q\ +\\
+\sum^3_{m=1}\sum^3_{q=1}
\hat T^i_m(\tau,\bry^1,\bry^2,\bry^3)\ \frac{\partial v^q}{\partial y^c}
\ \bar T^m_q
+\sum^3_{m=1}\sum^3_{q=1}
\hat T^i_m(\tau,\bry^1,\bry^2,\bry^3)\ v^q\ \frac{\partial
\bar T^m_q}{\partial y^c}.
\endgathered\quad
\mytag{3.28}
$$
Looking at the right hand side of \mythetag{3.28}, one easily recognizes
the partial derivative of the product of three terms. Therefore, 
\mythetag{3.28} is written as
$$
\pagebreak
\frac{\partial w^i}{\partial y^c}=\frac{\partial}{\partial y^c}
\!\left(\,\shave{\sum^3_{m=1}\sum^3_{q=1}}
\hat T^i_m(\tau,\bry^1,\bry^2,\bry^3)\ v^{\,q}\ \bar T^m_q\!\right)\!.
$$
Now, if we remember the formula \mythetag{3.12}, we obtain an even more
simple equality:
$$
\hskip -2em
\frac{\partial w^i}{\partial y^c}=\frac{\partial}{\partial y^c}
\!\left(\,\shave{\sum^3_{q=1}}v^{\,q}\ \hat T^i_q\right)\!.
\mytag{3.29}
$$
Integrating the equality \mythetag{3.29}, we define the components of
the vector $\bold w$:
$$
\hskip -2em
w^i=\sum^3_{q=1}v^{\,q}\ \hat T^i_q.
\mytag{3.30}
$$
This definition is unique up to some inessential terms depending 
only on the time variable: $w^i\to w^i+\omega^i(t)$. These terms can 
be omitted since $\bold w$ is used only in the form of its gradient
(see the equation \mythetag{1.4}).
\proclaim{Theorem 3.4}In the case of frozen dislocation the vectorial
parameter $\bold w$ is determined by the formula \mythetag{3.30}.
\endproclaim
\proclaim{Theorem 3.5}In the case of frozen dislocation the time evolution
of \ $\hat\bold T$, $\boldsymbol\rho$, $\bold j$, and $\bold w$ is determined
by the formulas \mythetag{3.12}, \mythetag{3.13} , \mythetag{3.24}, and \mythetag{3.30} respectively. Due to these formulas both differential
equations \mythetag{1.3} and \mythetag{1.4} are fulfilled identically.
\endproclaim
\par
\head
4. The analogy to electromagnetism and 
the gauge transformations.
\endhead
     The basic differential equations \mythetag{1.3} and \mythetag{1.4}
describing the kinematics of dislocations in crystals are similar
to the equations expressing the electric and magnetic fields $\bold E$ 
and $\bold H$ through the scalar potential $\varphi$ and the 
vector potential $\bold A$:
$$
\xalignat 2
&\hskip -2em
\bold H=\rot\bold A,
&&\bold E=-\grad\varphi-\frac{1}{c}\,\frac{\partial\bold A}{\partial t}.
\mytag{4.1}
\endxalignat
$$
These differential equations \mythetag{4.1} admit the following
gauge transformations
$$
\xalignat 2
&\hskip -2em
\bold A\to\bold A+\grad\psi,
&&\varphi\to\varphi-\frac{1}{c}\,\frac{\partial\psi}{\partial t}
\mytag{4.2}
\endxalignat
$$
that change potentials, but do not change the actual physical fields
$\bold E$ and $\bold H$ (see \mycite{3} for the reference). By analogy
we can write the gauge transformations
$$
\xalignat 2
&\hskip -2em
\hat\bold T \to\hat\bold T+\grad\boldsymbol\psi,
&&\bold w\to\bold w-\frac{\partial\boldsymbol\psi}{\partial t}
\mytag{4.3}
\endxalignat
$$
for $\hat\bold T$ and $\bold w$ in the differential equations
\mythetag{1.3} and \mythetag{1.4}. Despite to the striking similarity
of \mythetag{4.2} and \mythetag{4.3} their roles are absolutely 
different. The matter is that $\bold A$ and $\varphi$ in the 
electromagnetism are not actual physical fields, they are derived
mathematically from the Maxwell equations (see \mycite{3}). Unlike
$\bold A$, the incompatible distorsion field $\hat\bold T$ is 
an actual physical field describing the state of a medium. Applying 
the gauge transformation \mythetag{4.3} to it, we get another field 
describing some other state of the medium. For this reason, the 
nonlinear theory of dislocations is not a gauge theory or, at least, 
it is a gauge theory with the different gauge group, other than 
\pagebreak \mythetag{4.3}.\par
     The physical nature of the parameter $\bold w$ is still misty.
The following conjecture opens a way to clarify it. It is based on
the theorem~3.4.
\proclaim{Conjecture 4.1} The vectorial parameter $\bold w$ in the
equation \mythetag{1.4} is always determined by the formula \mythetag{3.30}.
\endproclaim
     If we admit the conjecture~4.1, then the differential equation
\mythetag{1.4} describing the time evolution of the incompatible distortion
tensor $\hat\bold T$ can be written as
$$
\hskip -2em
\frac{\partial\hat T^i_k}{\partial t}+
\sum^3_{p=1}\nabla_{\!p}\,\hat T^{p\,i}_k=-j^{\,i}_{\,k},
\mytag{4.4}
$$
where $\nabla_{\!p}=\partial/\partial y^p$ since we use the
Cartesian coordinates $y^1,\,y^2,\,y^3$, and
$$
\hskip -2em
\hat T^{p\,i}_k=\sum^3_{q=1}v^{\,q}\,\hat T^i_q\ \delta^p_k.
\mytag{4.5}
$$
In this form \mythetag{4.4} the differential equation \mythetag{1.4} is
quite similar to the balance equations which are traditional in the
mechanics of continuous media (see the mass balance equation, the
momentum balance equation, and the energy balance equation in \mycite{4}
for comparison). The tensor \mythetag{4.5} is interpreted as the density
of the incompatible distorsion flow.\par
     A conjecture is not a theorem yet. Therefore, we need to discuss
all of the possible options for the parameter $\bold w$ in the
equation \mythetag{1.4}:
\roster
\rosteritemwd=5pt
\item the parameter $\bold w$ is determined by the formula \mythetag{3.30};
\item the parameter $\bold w$ is determined by the formula
$$
\hskip -2em
w^i=\sum^3_{q=1}v^{\,q}\ T^i_q=-\frac{\partial T^i_q}{\partial t};
\mytag{4.6}
$$
\item the parameter $\bold w$ is determined by some formula other than
\mythetag{3.30} and \mythetag{4.6};
\item the parameter $\bold w$ is an independent parameter of a 
dislocated medium.
\endroster
\par
    The option \therosteritem{1} is my favorite option. It is supported
by the above two examples considered in the sections 2 and 3.\par
    The option \therosteritem{2} is often chosen in many papers. 
Substituting \mythetag{4.6} into the equation \mythetag{1.4}, we obtain
the following equality:
$$
\hskip -2em
\frac{\partial\bold T}{\partial t}=\frac{\partial\hat\bold T}{\partial t}
+\bold j\,.
\mytag{4.7}
$$    
The equality \mythetag{4.7} is referred to as the {\it additive decomposition} 
of the total distorsion into the {\it elastic\/} and {\it plastic\/} parts.
It is consistent in the linear theory, where both $\bold T$ and 
$\hat\bold T$ are approximately equal to the unit matrix. In the nonlinear
case the equality \mythetag{4.6} is not a valid option since it contradicts
the theorem~3.4. The {\it multiplicative decomposition\/} of the total deformation tensor suggested in \mycite{4} is more preferable.\par
    The choice of the option \therosteritem{3} or the option \therosteritem{4} produces more questions than the answers. One should
carry out a special research in order to exclude both these options or 
to prove one of them.
\head
5. Acknowledgments.
\endhead
    I am grateful to S.~F.~Lyuksyutov for the kind invitation to visit
Akron and for the opportunity to participate in the research in the field
of modern condensed matter physics. I am also grateful to the National Research Council and the National Academies of the USA for the financial
support of my visit through S.~F.~Lyuksyutov's COBASE grant.


\Refs
\ref\myrefno{1}
\by Comer~J., Sharipov~R.~A.\paper A note on the kinematics
of dislocations in crystals\publ e-print \myhref{http://arXiv.org/abs/math-ph/0410006/}{math-}\linebreak
\myhref{http://arXiv.org/abs/math-ph/0410006/}{ph/0410006}
in Electronic Archive \myEarXivlink
\endref
\ref\myrefno{2}
\by Sharipov~R.~A.\book Quick introduction to tensor analysis
\publ free on-line textbook in Electronic Archive \myEarXivlink;
see \myhref{http://arXiv.org/abs/math.HO/0403252}{math.HO/0403252}
\endref
\ref\myrefno{3}
\by Sharipov~R.~A.\book Classical electrodynamics and
theory of relativity\publ Bashkir State University\publaddr
Ufa, Russia\yr 1997\moreref English tr\.\yr 2003,
\myhref{http://arXiv.org/abs/physics/0311011/}{physics/0311011}
in Electronic Archive \myEarXivlink
\endref
\ref\myrefno{4}
\by Lyuksyutov~S.~F., Sharipov~R.~A.\paper Note on kinematics,
dynamics, and thermodynamics of plastic glassy media
\publ e-print \myhref{http://arXiv.org/abs/cond-mat/0304190/}
{cond-mat/0304190} in Electronic Archive \myEarXivlink
\endref
\endRefs
\enddocument
\end